\documentclass[aps,prl,reprint,superscriptaddress,nofootinbib,floatfix]{revtex4-2}
\usepackage{amsmath,amssymb,bm,booktabs,xcolor}
\usepackage{ytableau}
\usepackage{tikz}
\usetikzlibrary{arrows.meta}
\ytableausetup{boxsize=1.15em,centertableaux}
\usepackage[colorlinks=true,linkcolor=blue,citecolor=blue,urlcolor=blue]{hyperref}

\newif\ifcleanfinal
\cleanfinalfalse

\newcommand{\NF}{\operatorname{NF}}
\newcommand{\LM}{\operatorname{LM}}
\newcommand{\an}[2]{\langle #1\,#2\rangle}
\newcommand{\sq}[2]{[#1\,#2]}
\newcommand{\pp}[3]{\langle #1|p_{#2}|#3]}

\begin{document}

\title{Reducing Massive Scattering Amplitudes to Normal Form}
\author{Ming-Lei Xiao}
\affiliation{School of Science, Sun Yat-sen University Shenzhen Campus, Shenzhen 518107, P. R. China}

\begin{abstract}
	Local amplitude reduction is a crucial tool in effective field theory (EFT) calculations. While polynomial normal forms reproduce the massless y-basis, massive amplitudes require reduction rules that preserve little-group tensor contractions. We generalize Gr\"obner-basis reduction to covariant tensor identities, determining leading monomials in their contraction contexts and generating the higher-degree rules induced by mass determinants. Using three-, four-, and five-point replacement catalogs, we demonstrate symbolic-mass amplitude reductions and reproduce the independent three- and four-vector operator bases, providing traceable normal-form reductions for local EFT matching and syzygy analysis.
\end{abstract}
\maketitle

\indent\textit{Introduction.\enspace}
On-shell methods exploit the symmetries and analytic properties of scattering amplitudes to study particle interactions directly in terms of physical states~\cite{bddk,bcfw,elvanghuang}. Their extension to massive particles has opened a broad range of applications, from electroweak interactions to the gravitational dynamics of spinning bodies~\cite{allmassesspins,electroweak,spinningblackholes}. The massive spinor-helicity formalism provides a common language for these problems and has stimulated systematic constructions of amplitude bases for particles with arbitrary masses and spins~\cite{allmassesspins,massivecontact,dongshort,dong,deangelis}. Such a basis corresponds to the minimal set of local operators in effective field theories (EFTs), which provides the crucial input for calculations such as matching and renormalization-group (RG) running. However, the basis does not by itself give the coefficients of a given amplitude; determining them requires a standard procedure for expressing local amplitudes in that basis. Kinematic relations such as momentum conservation and Schouten identities connect algebraically different expressions of the amplitude, and can make the coefficients difficult to identify. For massless amplitudes, this reduction is achieved by the Young tensor method~\cite{ybasiseight,ybasisnine,li}, whose replacement rules admit a Gr\"obner-basis (GB) description~\cite{varieties}.

The Young tensor method was extended to massive particles in attempts to construct amplitude and operator bases for arbitrary masses and spins~\cite{dongshort,dong}, with subsequent developments using polarization configurations and high-energy limits~\cite{darkbasis}. These efforts have produced useful basis constructions, but reducing a general mass-dependent amplitude to the selected representatives remains a separate task. Exact on-shell sampling provides a route to coefficient extraction and basis conversion~\cite{efficientmatching,mosca}. Given a chosen spanning set, it numerically determines the coefficients of an amplitude in that set. In contrast, our goal is a symbolic procedure that reduces the input massive amplitude without first specifying a target spanning set.

GBs provide such an algebraic reduction and have already been used to remove kinematic redundancies in numerator ans\"atze for amplitude reconstruction~\cite{delaurentis,hh3j,hhh,qttH}. Massive applications employ composite generators, including momentum sandwiches and traces, whose relations are obtained by eliminating underlying spinor and momentum components; completing these relations remained unresolved in a massive-fermion example~\cite{qttH}. We develop a tensor Gr\"obner-basis construction for massive amplitudes, generalizing Buchberger's algorithm to covariant tensor identities. Leading monomials are determined in their contraction contexts, and every replacement strictly decreases the tensor monomial order. Completion resolves competing replacements through their $S$-polynomials; when all such ambiguities are resolved, the resulting tensor Gr\"obner basis yields unique normal forms. We demonstrate the method on three-, four-, and five-point amplitudes, generating the required higher-degree identities and reproducing the four-vector operator basis through four derivatives.

\par\smallskip
\indent\textit{Polynomial reduction and Gr\"obner bases.\enspace}
We start by illustrating the GB method for massless amplitude reduction in the polynomial \emph{ring} $R$ of spinor brackets. Kinematic constraints $r_a$ generate an \emph{ideal} $I\subset R$, and amplitude reduction seeks a distinguished representative, or \emph{normal form}, of each class in $R/I$ by repeatedly replacing monomials using these constraints until no rule applies. To make the replacements nonreversing, each replacement must decrease a \emph{monomial order} that respects multiplication and admits no infinite descent, ensuring termination. We use \emph{graded reverse lexicographic order}: compare bracket degrees first, then exponents from the smallest variable upward; at the first difference, the smaller exponent gives the larger monomial. For the spinor brackets used in this paper (with $A<B$ and $C<D$), the variable order is
\begin{equation}\label{eq:variable-order}
	\an{A}{B}\succ\an{C}{D}\succ\sq{C}{D}\succ\sq{A}{B}\,,
\end{equation}
with $(A,B)>_{\rm lex}(C,D)$, where $>_{\rm lex}$ compares the smaller label first, then the larger. The largest monomial in $r_a$ is its \emph{leading monomial} $\LM(r_a)$. For instance, in the Schouten identity $\mathsf S^{\rm a}_{1234}=\an{1}{4}\an{2}{3}-\an{1}{3}\an{2}{4}+\an{1}{2}\an{3}{4}$, the first term avoids the smallest brackets $\an{1}{2}$ and $\an{1}{3}$, so $\LM(\mathsf S^{\rm a}_{1234})=\an{1}{4}\an{2}{3}$, orienting the rule to replace it by the other two smaller terms.

\begin{figure}[tbp]
\centering
\begin{tikzpicture}[
	>=Stealth,
	monomial/.style={draw=black!70,fill=black!3,rounded corners=2pt,
		minimum width=11mm,minimum height=7mm,inner sep=2pt},
	reduction/.style={->,semithick},
	completion/.style={->,thick,densely dashed,draw=blue!65!black},
	rulelabel/.style={font=\small,fill=white,inner sep=2pt}
	]
	\node[monomial] (overlap) at (0,2.65) {$x^2y$};
	\node[monomial] (left) at (-1.45,-0.25) {$x$};
	\node[monomial] (right) at (1.45,0.85) {$y^2$};
	
	\draw[reduction] (overlap.south west)
	-- node[rulelabel,pos=0.48,left=2pt] {$r_2$} (left.north);
	\draw[reduction] (overlap.south east)
	-- node[rulelabel,pos=0.48,right=2pt] {$r_1$} (right.north);
	\draw[completion] (right.west)
	-- node[rulelabel,pos=0.2,below=7pt,text=blue!65!black] {$r_3=S(r_1,r_2)$} (left.east);
	
	\draw[->,semithick,draw=black!65] (-2.85,-0.65) -- (-2.85,3.05);
	\node[rotate=90,font=\small,text=black!75] at (-3.17,1.20)
	{Increasing monomial order};
\end{tikzpicture}
\caption{The input $x^2y$ ambiguously reduces to $x$ or $y^2$ by $\{r_1,r_2\}$. The $S$-polynomial supplies the third rule $r_3$, resolving the ambiguity. Height follows the monomial order $x^2y\succ y^2\succ x$.}
\label{fig:split-overlap}
\end{figure}
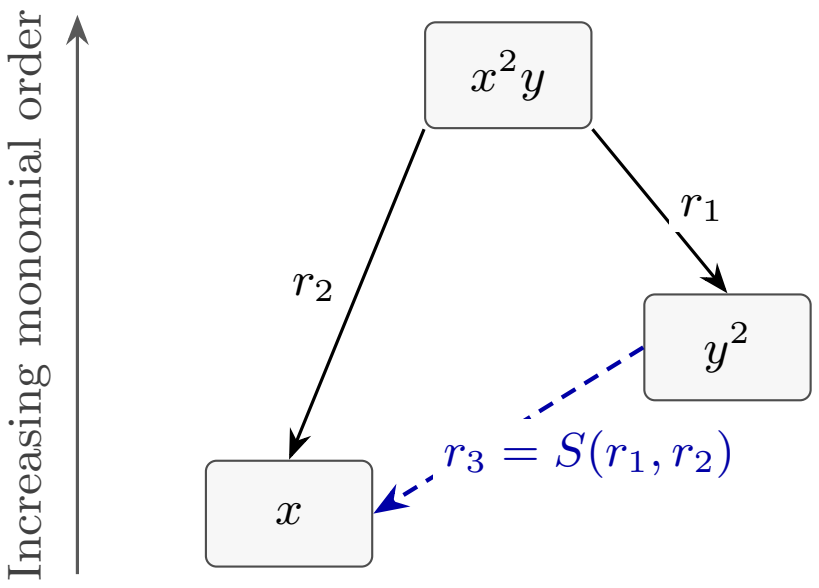

Termination alone does not guarantee a unique normal form: when two replacement rules apply to the same monomial, different choices may lead to different remainders. For two variables $x\succ y$ under this graded order, $r_1=x^2-y$ and $r_2=xy-1$ reduce $x^2y$ to $y^2$ and $x$, respectively. Cancelling the least common multiple of their LMs gives the $S$-polynomial
\begin{equation}
	S(r_1,r_2)=yr_1-xr_2=-y^2+x\equiv r_3.
	\label{uploaded:eq:split-overlap}
\end{equation}
Since $y^2\succ x$, the new rule $y^2\to x$ resolves the ambiguity as shown in Fig.~\ref{fig:split-overlap}, leaving the unique normal form $\NF(x^2y)=x$. Buchberger's algorithm adjoins nonzero reduced $S$-polynomials until all ambiguities resolve~\cite{coxLittleOShea}, yielding a GB $G=\{r_a\}$ of $I$ such that every polynomial in $I$ reduces to zero. Monomials not divisible by any $\LM(r_a)$ ($r_a\in G$) form the \emph{standard monomials} $\mathcal S$, spanning $R/I$. Reduction thus expresses any $f\in R$ as
\begin{equation}
	f=\sum_{r_a\in G}q_a r_a+\NF_G(f)\,,\quad
	\NF_G(f)=\sum_{m\in\mathcal S}c_m m\,, \label{uploaded:eq:split-normal}
\end{equation}
with its unique coordinate $c_m$ under the basis $\mathcal S$.

For massless amplitudes with $n\geq4$, momentum conservation $\sum_{k}\an{i}{k}\sq{k}{j}=0$ and both chiralities of Schouten identities generate the kinematic ideal. Under the monomial order~\eqref{eq:variable-order}, this ideal admits a complete quadratic GB~\cite{varieties} that reproduces the replacement rules of the Young tensor method~\cite{li} and requires no higher-degree rules. Its standard monomials precisely reproduce the Lorentz y-basis, represented by semistandard Young tableaux (SSYTs)~\cite{henning,ybasiseight,li}. For example, at $n=4$,
\begin{equation}
	\underbrace{
		\begin{ytableau}
			*(blue!15) 3 & 1 & 1\\
			*(blue!15) 4 & 2 & 3
		\end{ytableau}
	}_{\substack{\text{non-standard}\\[2pt] \sq{1}{2}\an{1}{2}\an{1}{3}}}
	\ \xrightarrow{\ \mathrm{GB}\ }\ 
	\underbrace{
		\begin{ytableau}
			*(blue!15) 1 & 1 & 3\\
			*(blue!15) 2 & 3 & 4
		\end{ytableau}
	}_{\substack{\text{SSYT}\\[2pt] \sq{3}{4}\an{1}{3}\an{3}{4}}} .
	\label{eq:split-massless-example}
\end{equation}
The blue columns represent square brackets through their complementary particle labels, while angle-bracket columns carry their labels directly. Momentum-conservation rules eliminate $\an{1}{2}\sq{1}{2}$ or $\an{1}{3}\sq{1}{2}$ in either order to yield the same semistandard tableau, with weakly increasing rows and strictly increasing columns.

\par\bigskip
\indent\textit{Massive amplitude reduction.---}
Massive amplitude reduction differs from the massless problem in two respects. First, imposing the mass-determinant relations $\det|\boldsymbol{i}\rangle=\det|\boldsymbol{i}]=m_i$~\cite{allmassesspins} introduces mass-dependent terms of lower bracket degree into the identities. These terms can survive the Buchberger procedure, leaving nonzero reduced $S$-polynomials that supply new replacement rules. Three massive vectors provide an example (with bold labels suppressing symmetrized free little-group indices on each leg): the eight monomials
\begin{equation}
	\begin{gathered}
		\an{\boldsymbol{1}}{\boldsymbol{2}}\an{\boldsymbol{2}}{\boldsymbol{3}}\an{\boldsymbol{3}}{\boldsymbol{1}},\quad \sq{\boldsymbol{1}}{\boldsymbol{2}}\sq{\boldsymbol{2}}{\boldsymbol{3}}\sq{\boldsymbol{3}}{\boldsymbol{1}},\\
		\sq{\boldsymbol{1}}{\boldsymbol{2}}\an{\boldsymbol{2}}{\boldsymbol{3}}\an{\boldsymbol{3}}{\boldsymbol{1}},\quad \an{\boldsymbol{1}}{\boldsymbol{2}}\sq{\boldsymbol{2}}{\boldsymbol{3}}\an{\boldsymbol{3}}{\boldsymbol{1}},\\
		\an{\boldsymbol{1}}{\boldsymbol{2}}\an{\boldsymbol{2}}{\boldsymbol{3}}\sq{\boldsymbol{3}}{\boldsymbol{1}},\quad \an{\boldsymbol{1}}{\boldsymbol{2}}\sq{\boldsymbol{2}}{\boldsymbol{3}}\sq{\boldsymbol{3}}{\boldsymbol{1}},\\
		\sq{\boldsymbol{1}}{\boldsymbol{2}}\an{\boldsymbol{2}}{\boldsymbol{3}}\sq{\boldsymbol{3}}{\boldsymbol{1}},\quad \sq{\boldsymbol{1}}{\boldsymbol{2}}\sq{\boldsymbol{2}}{\boldsymbol{3}}\an{\boldsymbol{3}}{\boldsymbol{1}}
	\end{gathered}
	\label{eq:three-vector-candidates}
\end{equation}
obey a hidden identity $\mathsf{F}=0$, where
\begin{equation}
	\mathsf{F}=\sum_{\rm cyc(1,2,3)}m_3\bigl(\sq{\boldsymbol{1}}{\boldsymbol{2}}\an{\boldsymbol{3}}{\boldsymbol{1}}\an{\boldsymbol{2}}{\boldsymbol{3}} + \an{\boldsymbol{1}}{\boldsymbol{2}}\sq{\boldsymbol{3}}{\boldsymbol{1}}\sq{\boldsymbol{2}}{\boldsymbol{3}}\bigr) .
	\label{eq:three-vector-F}
\end{equation}
None of these eight monomials is reducible by the original momentum or Schouten rules, yet $\mathsf{F}=0$---generated below as a reduced $S$-polynomial---relates six of them, leaving seven independent structures corresponding to the minimal set of seven on-shell $W^+W^-Z$ couplings~\cite{hagiwara1987,zeppenfeld1988}.

Second, unlike massless monomials, which are simple products of brackets, massive monomials also involve contractions of little-group indices between brackets. Their covariant tensor structure is therefore not captured by an ordinary polynomial ring of bracket variables, so Buchberger's algorithm cannot be applied directly.\footnote{The GB method can be applied to individual little-group components, but this need not yield covariant tensor normal forms. Alternatively, it can be applied to a polynomial ring of composite tensors with tensor-product multiplication~\cite{hh3j,qttH}.} 
We extend Buchberger's construction to tensor replacement rules by redefining divisibility and the ordering prescription to account for these contractions.

A tensor monomial $L$ \emph{divides} $T$ if $T$ can be obtained by multiplying $L$ by additional brackets with a specified pattern of little-group contractions, which together constitute the contraction context. For example, $\an{\boldsymbol{1}}{\boldsymbol{2}}$ divides both $\an{\boldsymbol{1}}{\boldsymbol{2}}\sq{\boldsymbol{2}}{\boldsymbol{3}}$ and $\pp{\boldsymbol{1}}{2}{\boldsymbol{3}}$, with the same additional bracket $\sq{\boldsymbol{2}}{\boldsymbol{3}}$ but different contractions. A replacement applies when the identity's LM divides the input monomial, oriented by the following successive comparisons:
\begin{enumerate}
	\setlength{\itemsep}{1pt}\setlength{\parsep}{0pt}
	\item \emph{Bracket degree:} Masses are treated as coefficients not counted in bracket degree, orienting identities toward terms with more mass factors, such as $\pp{\boldsymbol{1}}{2}{\boldsymbol{3}}\succ m_3\an{\boldsymbol{1}}{\boldsymbol{3}}$.\footnote{Alternatively, masses may be retained as polynomial variables of degree one, which is useful for nonrelativistic EFTs.}
	\item \emph{Bracket product:} At equal degree, disregard little-group indices and contractions (unbolding) and compare under the massless SSYT-compatible order~\eqref{eq:variable-order}, e.g., $\an{\boldsymbol{1}}{\boldsymbol{4}}\an{\boldsymbol{2}}{\boldsymbol{3}}\succ\an{\boldsymbol{1}}{\boldsymbol{3}}\an{\boldsymbol{2}}{\boldsymbol{4}}$, ensuring compatibility with massless reductions.
	\item \emph{Contraction structure:} If the unbolded bracket products coincide, a tie-breaker orders the contraction patterns. For example, ranking longer contraction chains higher breaks traces into shorter chains,
	\begin{equation}
		\operatorname{tr}(p_2\bar p_3p_2\bar p_4)
		\to \operatorname{tr}(p_2\bar p_3)\operatorname{tr}(p_2\bar p_4) - m_2^2\operatorname{tr}(p_3\bar p_4) \,.
	\end{equation}
	Full details of the tie-breaker are given in End Matter.
\end{enumerate}

\begin{figure}[tb]
\centering
\begin{tikzpicture}[font=\footnotesize,
 comparison/.style={draw,rounded corners=2pt,fill=white,inner sep=5pt},
 edge/.style={semithick}]
\node[comparison] (r) at (-2,0) {$T_1^{(r)}\succ T_2^{(r)}$};
\node[comparison] (s) at (0,2.8) {$T_1^{(s)}\prec T_2^{(s)}$};
\node[comparison] (t) at (2,0) {$T_1^{(t)}\succ T_2^{(t)}$};
\draw[edge] (r)--(s) node[pos=0.45,sloped,above,font=\scriptsize]
 {$D_1^{(r,s)}\succ D_2^{(r,s)}$};
\draw[edge] (s)--(t) node[pos=0.55,sloped,above,font=\scriptsize]
 {$D_1^{(s,t)}\prec D_2^{(s,t)}$};
\draw[edge,red] (t)--(r) node[midway,above,font=\scriptsize]
 {$D_1^{(t,r)}\mathrel{?}D_2^{(t,r)}$};
\end{tikzpicture}
\caption{Obstruction to context-preserving order. Each edge compares the monomials obtained by attaching the other spectator. Exchanging the attachment order interchanges $D_1$ and $D_2$, so order inheritance requires opposite comparisons at adjacent vertices. The first two edges force equal comparisons at $r$ and $t$, imposing contradictory orders on the third edge.}
\label{fig:context-order-triangle}
\end{figure}
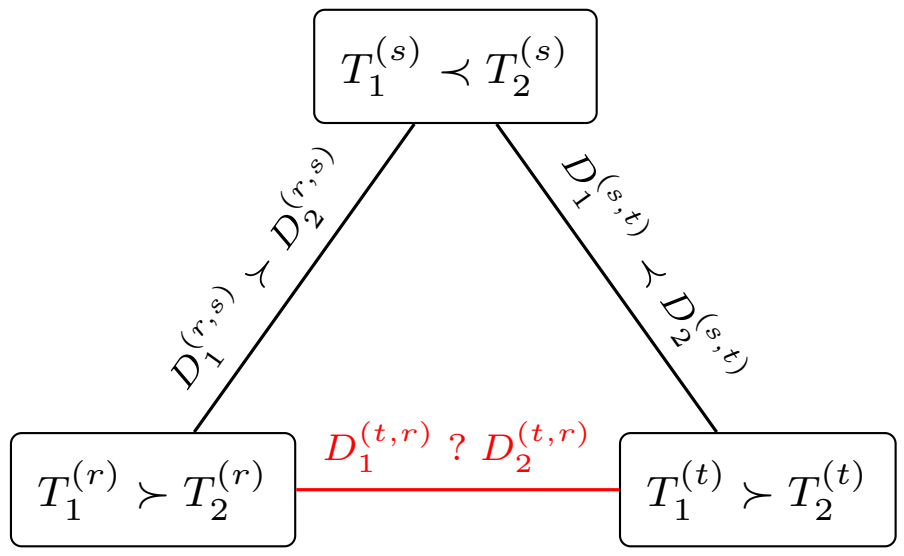

One important subtlety is that while the first two comparisons are preserved under allowed contractions, the third tie-breaker need not be. As a demonstration, take $n\geq6$ and choose the last momentum $p_n$---which is not reduced by momentum conservation---together with three distinct spectator labels $r,s,t\notin\{1,2,n\}$. The two monomials
\begin{align}
	T_1^{(r)}&=\an{\boldsymbol{1}}{\boldsymbol{n}}\pp{\boldsymbol{2}}{n}{\boldsymbol{r}},&
	T_2^{(r)}&=\an{\boldsymbol{2}}{\boldsymbol{n}}\pp{\boldsymbol{1}}{n}{\boldsymbol{r}}, \label{eq:tensor_T}
\end{align}
are related by the Schouten identity $T_1^{(r)}-T_2^{(r)}+m_n\an{\boldsymbol{1}}{\boldsymbol{2}}\sq{\boldsymbol{n}}{\boldsymbol{r}}=0$, whose LM must be chosen between $T_1^{(r)}$ and $T_2^{(r)}$ by their contraction structure. Attaching $\sq{\boldsymbol{n}}{\boldsymbol{s}}$ to the remaining free angle index on particle $n$ produces, respectively,
\begin{equation}
	D_1^{(r,s)}=\pp{\boldsymbol{1}}{n}{\boldsymbol{s}}\pp{\boldsymbol{2}}{n}{\boldsymbol{r}},\quad
	D_2^{(r,s)}=\pp{\boldsymbol{2}}{n}{\boldsymbol{s}}\pp{\boldsymbol{1}}{n}{\boldsymbol{r}}.
\end{equation}
If contractions preserved the order, the comparison of $D_1^{(r,s)}$ and $D_2^{(r,s)}$ would follow that of $T_1^{(r)}$ and $T_2^{(r)}$. Starting instead with spectator $s$ and attaching $\sq{\boldsymbol{n}}{\boldsymbol{r}}$ yields the same pair in reverse order, $(D_1^{(s,r)},D_2^{(s,r)})=(D_2^{(r,s)},D_1^{(r,s)})$. Order inheritance therefore requires opposite comparisons for spectators $r$ and $s$, which is impossible to satisfy simultaneously on all three pairs $(r,s)$, $(s,t)$, and $(t,r)$ (Fig.~\ref{fig:context-order-triangle}). The leading term must therefore be determined in context: every term of an identity sharing the leading unbolded product is retained as a potential LM, and their images are compared in the contraction context of the input amplitude.

Using this order, we build the complete rule set starting from the basic kinematic identities. For $i\ne j$, momentum conservation acquires mass terms,
\begin{equation}
	\mathsf M_{ij}=\sum_{k\ne i,j}\pp{\boldsymbol{i}}{k}{\boldsymbol{j}}
	-m_i\sq{\boldsymbol{i}}{\boldsymbol{j}}-m_j\an{\boldsymbol{i}}{\boldsymbol{j}}=0,
	\label{uploaded:eq:v3-M}
\end{equation}
whose LM has the smallest label $k\ne i,j$ ($k\in\{1,2,3\}$), alongside the diagonal ($i=j$) Mandelstam relations. Distinct-label Schouten identities ($i<j<k<l$) retain their massless form and first-term LM,
\begin{align}
	\mathsf S^{\rm a}_{ijkl}={}&\an{\boldsymbol{i}}{\boldsymbol{l}}\an{\boldsymbol{j}}{\boldsymbol{k}}
	-\an{\boldsymbol{i}}{\boldsymbol{k}}\an{\boldsymbol{j}}{\boldsymbol{l}} + \an{\boldsymbol{i}}{\boldsymbol{j}}\an{\boldsymbol{k}}{\boldsymbol{l}}=0,
\end{align}
and similarly for $\mathsf S^{\rm s}_{ijkl}$ with square brackets. Massive particles also obey the repeated-label Schouten identities (exposing the open indices $I,J$ and $K,L$)
\begin{align}
	\mathsf S^{\rm a}_{iijk}={}&\an{\boldsymbol{i}^I}{\boldsymbol{j}}\an{\boldsymbol{i}^J}{\boldsymbol{k}}
	-\an{\boldsymbol{i}^J}{\boldsymbol{j}}\an{\boldsymbol{i}^I}{\boldsymbol{k}}-m_i\epsilon^{IJ}\an{\boldsymbol{j}}{\boldsymbol{k}}=0,\\
	\mathsf S^{\rm a}_{iijj}={}&\an{\boldsymbol{i}^I}{\boldsymbol{j}^K}\an{\boldsymbol{i}^J}{\boldsymbol{j}^L}-\an{\boldsymbol{i}^I}{\boldsymbol{j}^L}\an{\boldsymbol{i}^J}{\boldsymbol{j}^K}\nonumber\\
	&-m_i m_j\epsilon^{IJ}\epsilon^{KL}=0,
\end{align}
and their square-bracket counterparts $\mathsf S^{\rm s}_{iijk},\mathsf S^{\rm s}_{iijj}$. In each repeated-label identity, the two quadratic terms share the same unbolded product while the mass term has lower degree; both quadratic terms are kept as potential LMs and oriented in context, as in Eq.~\eqref{eq:tensor_T}.

As in Buchberger's algorithm, overlapping replacements can give different reduction routes for the same monomial. Consider the three-vector monomial
\begin{equation}
	H=\an{\boldsymbol{3}}{\boldsymbol{1}}\pp{\boldsymbol{2}}{3}{\boldsymbol{2}}
	\sq{\boldsymbol{3}}{\boldsymbol{1}}.
\end{equation}
The repeated-label Schouten identities $\mathsf S^{\rm a}_{1233}$ and $\mathsf S^{\rm s}_{1233}$ act on its angle and square brackets, respectively, giving two routes:
\begin{align}
	H\xrightarrow{\mathsf S^{\rm s}_{1233}}{}&
	-\an{\boldsymbol{3}}{\boldsymbol{1}}\pp{\boldsymbol{2}}{3}{\boldsymbol{1}}\sq{\boldsymbol{2}}{\boldsymbol{3}}
	\nonumber\\[-2pt]
	&+m_3\sq{\boldsymbol{1}}{\boldsymbol{2}}\an{\boldsymbol{3}}{\boldsymbol{1}}\an{\boldsymbol{2}}{\boldsymbol{3}}
	\xrightarrow{\mathsf M_{21}}R_{\rm s},\\
	H\xrightarrow{\mathsf S^{\rm a}_{1233}}{}&
	-\an{\boldsymbol{2}}{\boldsymbol{3}}\pp{\boldsymbol{1}}{3}{\boldsymbol{2}}\sq{\boldsymbol{3}}{\boldsymbol{1}}
	\nonumber\\[-2pt]
	&-m_3\an{\boldsymbol{1}}{\boldsymbol{2}}\sq{\boldsymbol{3}}{\boldsymbol{1}}\sq{\boldsymbol{2}}{\boldsymbol{3}}
	\xrightarrow{\mathsf M_{12}}R_{\rm a},
\end{align}
where eliminating $\pp{\boldsymbol{2}}{3}{\boldsymbol{1}}$ and $\pp{\boldsymbol{1}}{3}{\boldsymbol{2}}$ via $\mathsf M_{12}$ and $\mathsf M_{21}$ yields
\begin{align}
	R_{\rm s}&=m_1\an{\boldsymbol{1}}{\boldsymbol{2}}\an{\boldsymbol{3}}{\boldsymbol{1}}\sq{\boldsymbol{2}}{\boldsymbol{3}} + m_2\sq{\boldsymbol{1}}{\boldsymbol{2}}\an{\boldsymbol{3}}{\boldsymbol{1}}\sq{\boldsymbol{2}}{\boldsymbol{3}} \nonumber\\
	&\qquad + m_3\sq{\boldsymbol{1}}{\boldsymbol{2}}\an{\boldsymbol{3}}{\boldsymbol{1}}\an{\boldsymbol{2}}{\boldsymbol{3}},\\
	R_{\rm a}&=-m_1\sq{\boldsymbol{1}}{\boldsymbol{2}}\sq{\boldsymbol{3}}{\boldsymbol{1}}\an{\boldsymbol{2}}{\boldsymbol{3}} - m_2\an{\boldsymbol{1}}{\boldsymbol{2}}\sq{\boldsymbol{3}}{\boldsymbol{1}}\an{\boldsymbol{2}}{\boldsymbol{3}} \nonumber\\
	&\qquad - m_3\an{\boldsymbol{1}}{\boldsymbol{2}}\sq{\boldsymbol{3}}{\boldsymbol{1}}\sq{\boldsymbol{2}}{\boldsymbol{3}}.
\end{align}
Subtracting the two contextual Schouten identities cancels $H$, and reducing their difference by momentum conservation gives
\begin{equation}
	S\bigl(\mathsf S^{\rm a}_{1233}, \mathsf S^{\rm s}_{1233}\bigr) \xrightarrow{\mathsf M_{12},\,\mathsf M_{21}} R_{\rm s}-R_{\rm a} = \mathsf F ,
\end{equation}
recovering $\mathsf F$ in Eq.~\eqref{eq:three-vector-F} and supplying the missing degree-three rule among the eight candidates in Eq.~\eqref{eq:three-vector-candidates}.

To find such overlaps systematically, we construct a \emph{minimal common contraction} of two potential LMs. Here the angle head $\an{\boldsymbol{3}}{\boldsymbol{1}}\an{\boldsymbol{2}}{\boldsymbol{3}}$ and square head $\sq{\boldsymbol{3}}{\boldsymbol{2}}\sq{\boldsymbol{3}}{\boldsymbol{1}}$ are joined into $H$ by contracting the particle-3 indices on $\an{\boldsymbol{2}}{\boldsymbol{3}}$ and $\sq{\boldsymbol{3}}{\boldsymbol{2}}$, making $H$ the contextual LM of both identities. An ordinary polynomial LCM would retain only the four uncontracted bracket factors, losing the contraction that defines the overlap. While the LCMs in Refs.~\cite{hh3j,qttH} are defined through tensor products, our minimal common contractions also allow tensor contractions.

\begin{figure}[tb]
	\centering
	\begin{tikzpicture}[>=Stealth,font=\scriptsize,
		box/.style={draw,rounded corners=2pt,align=center,text width=30mm,minimum height=7mm,inner sep=3pt},
		arr/.style={->,semithick}]
		\node[font=\footnotesize\bfseries] at (-1.9,0.6) {Construction};
		\node[font=\footnotesize\bfseries] at (1.9,-0.5) {Amplitude reduction};
		\node[box] (c1) at (-1.9,0) {Momentum + Schouten};
		\node[box] (c2) at (-1.9,-1.1) {Common contractions\\Check contextual LMs};
		\node[box] (c3) at (-1.9,-2.2) {Form and reduce\\$S$-polynomials};
		\node[box] (c4) at (-1.9,-3.3) {Add new rules};
		\node[box] (c5) at (-1.9,-4.4) {Replacement rules};
		\draw[arr] (c1)--(c2);\draw[arr] (c2)--(c3);
		\draw[arr] (c3)--(c4);\draw[arr] (c4)--(c5);
		\draw[arr] (c4.west)--++(-0.25,0)|-(c2.west);
		\node[box] (r1) at (1.9,-1.1) {Input amplitude\\Highest degree first};
		\node[box] (r2) at (1.9,-2.2) {Match context\\Determine LM};
		\node[box,text width=14mm] (r3) at (2.70,-3.50) {Apply\\reduction};
		\node[box] (r5) at (1.9,-4.4) {Normal form};
		\draw[arr] (r1)--(r2);
		\draw[arr] (r2.south -| r3.north)--node[left,font=\tiny,align=right] {rule\\applies}(r3.north);
		\draw[arr] (r2.south -| 0.70,0)--node[right,font=\tiny,align=left] {no rule\\applies}(r5.north -| 0.70,0);
		\draw[arr] (r3.east)--++(0.25,0)|-(r2.east);
		\draw[arr] (c5.east)--(-0.1,-4.4)--(-0.1,-2.2)--(r2.west);
	\end{tikzpicture}
	\caption{Workflow for tensor rule construction (left) and top-down amplitude reduction to normal form (right).}
	\label{fig:algorithm-workflow}
\end{figure}
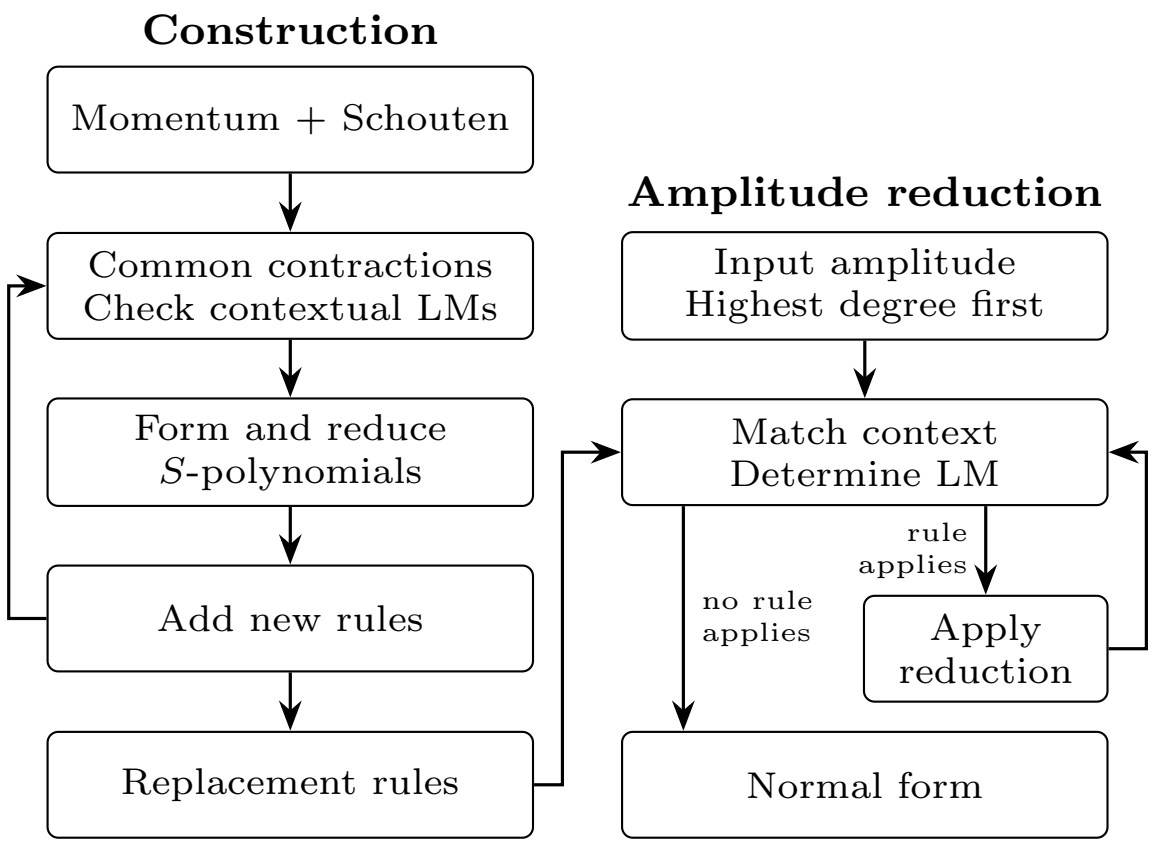

Figure~\ref{fig:algorithm-workflow} summarizes rule construction and amplitude reduction. Starting from the momentum and Schouten identities, we enumerate minimal common contractions of potential LMs, verify that the shared monomial is leading in both contextual identities, and reduce their $S$-polynomial; each nonzero remainder is adjoined as a new rule to form further overlaps. Amplitude reduction then proceeds from the highest bracket degree downward, repeatedly applying rules in matched contraction contexts to strictly decrease the monomial order until no rule applies, yielding the normal form.

\par\smallskip
\indent\textit{Implementation and Results.\enspace}
We implement the algorithm to construct tensor GBs for $n$-point massive amplitudes. To constrain the range of search for non-zero reduced $S$-polynomials as new rules, we impose a bound for the bracket degree $d$ of their LM and obtain the rule catalog $(n,d)$. Table~\ref{tab:rule-degree-counts} lists the rules by degree for $n=3,4,5$. At three points, the rule set saturates at $d=4$, consistent with the absence of independent Mandelstam variables. Interestingly, at four points, our search through $d=8$ found no new rules beyond $d=6$, also suggesting saturation. However, we found no sign of saturation at five points. The catalogs $(3,4)$, $(4,6)$ and $(5,6)$, Python construction code, and Mathematica sample code are available in the \href{https://github.com/XiaoMing-lei/MassiveAmplitudeReduction}{public code repository}.
\begin{table}[tb]
\centering\small
\setlength{\tabcolsep}{4pt}
\begin{tabular}{crrrrrrr}
\toprule
$n$ & $d=2$ & $3$ & $4$ & $5$ & $6$ & $7$ & Total\\
\midrule
3 &22&4&1&---&---&---&27\\
4 &55&12&67&38&7&---&179\\
5 &116&23&299&279&360&\multicolumn{2}{c}{N/A}\\
\bottomrule
\end{tabular}
\caption{Rule counts by bracket degree for $n=3,4,5$ massive amplitudes, including both the original momentum and Schouten rules and the generated rules. The total column reports the observed saturated count; N/A at $n=5$ indicates that the rule count has not saturated within the explored range.}
\label{tab:rule-degree-counts}
\end{table}

With these catalogs, we provide Mathematica sample code for reducing massive amplitudes and recording the applied rules. We illustrate it with three benchmarks,
\begin{align}
\mathcal A_1(V_1,V_2,V_3,V_4)&=\an{\boldsymbol1}{\boldsymbol2}\sq{\boldsymbol2}{\boldsymbol3}\sq{\boldsymbol3}{\boldsymbol4}
\langle\boldsymbol1|p_3\bar p_2|\boldsymbol4\rangle,\nonumber\\
\mathcal A_2(\psi_1,V_2,V_3,\psi_4)&=\an{\boldsymbol1}{\boldsymbol4}\an{\boldsymbol2}{\boldsymbol3}^{2}
\operatorname{tr}\bigl(p_1\bar{p}_2p_3\bar{p}_4\bigr),\nonumber\\
\mathcal A_3(V_1,V_2,V_3,V_4,\phi_5)&=\an{\boldsymbol1}{\boldsymbol2}\an{\boldsymbol3}{\boldsymbol4}
\sq{\boldsymbol1}{\boldsymbol3}\sq{\boldsymbol2}{\boldsymbol4}\operatorname{tr}\bigl(p_1\bar{p}_5\bigr),
\end{align}
where $V$, $\psi$ and $\phi$ denote vector, fermion and scalar particles. All spinor indices inside the trace are contracted. Every input factor participates in the selected reduction sequence. The four-point examples use the $(4,6)$ catalog, and the five-point example uses $(5,6)$. Each resulting normal form is verified by step-by-step symbolic replay and independent numerical on-shell evaluations; Table~\ref{tab:implementation-benchmarks} reports the reduction statistics.
\begin{table}[htb]
\centering\small
\setlength{\tabcolsep}{6pt}
\begin{tabular}{crrr}
\toprule
Input & $d$ & $N_{\rm out}$ & $N_{\rm step}$\\
\midrule
$\mathcal A_1$&6&15&14\\
$\mathcal A_2$&7&22&24\\
$\mathcal A_3$&6&31&37\\
\bottomrule
\end{tabular}
\caption{Four- and five-point reductions. $N_{\rm out}$ counts collected tensor monomials in the normal form; $N_{\rm step}$ counts successive rule applications along the selected reduction sequence.}
\label{tab:implementation-benchmarks}
\end{table}
The notebook reports reduction times separately from catalog loading; its supplied prepared replacements can be reused for repeated inputs. The notebook also presents the detailed steps of reduction, such as the $N_{\rm step}=14$ successive rule applications for $\mathcal A_1$. Contextual replacements derived from the catalogs can also be cached to accelerate subsequent reductions.

As an independent check, we construct the minimal basis of four-distinguishable-massive-vector amplitudes corresponding to the operators $V^4D^0$, $V^4D^2$ and $V^4D^4$ by reducing all 39,945 candidate monomials. The scan found 4, 56 and 141 irreducible monomials, respectively, precisely corresponding to the independent bases of effective operators at dimension four, six and eight. The numbers are in agreement with existing constructions~\cite{dong,darkbasis,massiveamplitudecode}; the four-derivative comparison uses the authors' code. While more computationally intensive than direct basis enumeration, this scan simultaneously expresses all the candidate monomials as explicit linear combinations of the 201 basis amplitudes. Exact evaluations at on-shell kinematic points give a matrix of rank 201, confirming the linear independence of the cumulative basis. 

The tensor Gr\"obner-basis approach reduces local massive amplitudes to normal form, supporting EFT matching and RG calculations by expressing local amplitudes and operator matrix elements with the same rules. Even without constructing a full sector basis, reducing any subset of massive amplitudes to normal form suffices to establish their independence or determine the syzygies (dependency relations) among them. A natural next step is to extend the method to mixed massless-massive amplitudes, bringing a wider range of physical processes into a unified reduction framework. Further directions include applying the reduction at the operator level for Lagrangian calculations and adapting the construction to rational functions to systematically simplify loop-level amplitude expressions.

\par\medspace
\begin{acknowledgments}
\noindent\textit{Acknowledgments.---}
M.-L.X. is supported by the National Natural Science Foundation of China under Grant No.~12405123, by the Fundamental Research Funds for the Central Universities at Sun Yat-Sen University under Grant No.~25hytd001, and by the Shenzhen Science and Technology Program under Grant No.~JCYJ20240813150911015. OpenAI Codex was used to assist with algorithm implementation. The author takes responsibility for the scientific content and conclusions.
\end{acknowledgments}
\bibliographystyle{apsrev4-2}
\bibliography{references}

\begin{thebibliography}{26}%
\makeatletter
\providecommand \@ifxundefined [1]{%
 \@ifx{#1\undefined}
}%
\providecommand \@ifnum [1]{%
 \ifnum #1\expandafter \@firstoftwo
 \else \expandafter \@secondoftwo
 \fi
}%
\providecommand \@ifx [1]{%
 \ifx #1\expandafter \@firstoftwo
 \else \expandafter \@secondoftwo
 \fi
}%
\providecommand \natexlab [1]{#1}%
\providecommand \enquote  [1]{``#1''}%
\providecommand \bibnamefont  [1]{#1}%
\providecommand \bibfnamefont [1]{#1}%
\providecommand \citenamefont [1]{#1}%
\providecommand \href@noop [0]{\@secondoftwo}%
\providecommand \href [0]{\begingroup \@sanitize@url \@href}%
\providecommand \@href[1]{\@@startlink{#1}\@@href}%
\providecommand \@@href[1]{\endgroup#1\@@endlink}%
\providecommand \@sanitize@url [0]{\catcode `\\12\catcode `\$12\catcode
  `\&12\catcode `\#12\catcode `\^12\catcode `\_12\catcode `\%12\relax}%
\providecommand \@@startlink[1]{}%
\providecommand \@@endlink[0]{}%
\providecommand \url  [0]{\begingroup\@sanitize@url \@url }%
\providecommand \@url [1]{\endgroup\@href {#1}{\urlprefix }}%
\providecommand \urlprefix  [0]{URL }%
\providecommand \Eprint [0]{\href }%
\providecommand \doibase [0]{https://doi.org/}%
\providecommand \selectlanguage [0]{\@gobble}%
\providecommand \bibinfo  [0]{\@secondoftwo}%
\providecommand \bibfield  [0]{\@secondoftwo}%
\providecommand \translation [1]{[#1]}%
\providecommand \BibitemOpen [0]{}%
\providecommand \bibitemStop [0]{}%
\providecommand \bibitemNoStop [0]{.\EOS\space}%
\providecommand \EOS [0]{\spacefactor3000\relax}%
\providecommand \BibitemShut  [1]{\csname bibitem#1\endcsname}%
\let\auto@bib@innerbib\@empty
\bibitem [{\citenamefont {Bern}\ \emph {et~al.}(1994)\citenamefont {Bern},
  \citenamefont {Dixon}, \citenamefont {Dunbar},\ and\ \citenamefont
  {Kosower}}]{bddk}%
  \BibitemOpen
  \bibfield  {author} {\bibinfo {author} {\bibfnamefont {Z.}~\bibnamefont
  {Bern}}, \bibinfo {author} {\bibfnamefont {L.}~\bibnamefont {Dixon}},
  \bibinfo {author} {\bibfnamefont {D.~C.}\ \bibnamefont {Dunbar}},\ and\
  \bibinfo {author} {\bibfnamefont {D.~A.}\ \bibnamefont {Kosower}},\ }\href
  {https://doi.org/10.1016/0550-3213(94)90179-1} {\bibfield  {journal}
  {\bibinfo  {journal} {Nucl. Phys. B}\ }\textbf {\bibinfo {volume} {425}},\
  \bibinfo {pages} {217} (\bibinfo {year} {1994})},\ \Eprint
  {https://arxiv.org/abs/hep-ph/9403226} {arXiv:hep-ph/9403226} \BibitemShut
  {NoStop}%
\bibitem [{\citenamefont {Britto}\ \emph {et~al.}(2005)\citenamefont {Britto},
  \citenamefont {Cachazo}, \citenamefont {Feng},\ and\ \citenamefont
  {Witten}}]{bcfw}%
  \BibitemOpen
  \bibfield  {author} {\bibinfo {author} {\bibfnamefont {R.}~\bibnamefont
  {Britto}}, \bibinfo {author} {\bibfnamefont {F.}~\bibnamefont {Cachazo}},
  \bibinfo {author} {\bibfnamefont {B.}~\bibnamefont {Feng}},\ and\ \bibinfo
  {author} {\bibfnamefont {E.}~\bibnamefont {Witten}},\ }\href
  {https://doi.org/10.1103/PhysRevLett.94.181602} {\bibfield  {journal}
  {\bibinfo  {journal} {Phys. Rev. Lett.}\ }\textbf {\bibinfo {volume} {94}},\
  \bibinfo {pages} {181602} (\bibinfo {year} {2005})},\ \Eprint
  {https://arxiv.org/abs/hep-th/0501052} {arXiv:hep-th/0501052} \BibitemShut
  {NoStop}%
\bibitem [{\citenamefont {Elvang}\ and\ \citenamefont
  {Huang}(2013)}]{elvanghuang}%
  \BibitemOpen
  \bibfield  {author} {\bibinfo {author} {\bibfnamefont {H.}~\bibnamefont
  {Elvang}}\ and\ \bibinfo {author} {\bibfnamefont {Y.-t.}\ \bibnamefont
  {Huang}},\ }\href@noop {} {\bibinfo {title} {Scattering amplitudes}}
  (\bibinfo {year} {2013}),\ \Eprint {https://arxiv.org/abs/1308.1697}
  {arXiv:1308.1697 [hep-th]} \BibitemShut {NoStop}%
\bibitem [{\citenamefont {Arkani-Hamed}\ \emph {et~al.}(2017)\citenamefont
  {Arkani-Hamed}, \citenamefont {Huang},\ and\ \citenamefont
  {Huang}}]{allmassesspins}%
  \BibitemOpen
  \bibfield  {author} {\bibinfo {author} {\bibfnamefont {N.}~\bibnamefont
  {Arkani-Hamed}}, \bibinfo {author} {\bibfnamefont {T.-C.}\ \bibnamefont
  {Huang}},\ and\ \bibinfo {author} {\bibfnamefont {Y.-t.}\ \bibnamefont
  {Huang}},\ }\href@noop {} {\bibinfo {title} {Scattering amplitudes for all
  masses and spins}} (\bibinfo {year} {2017}),\ \Eprint
  {https://arxiv.org/abs/1709.04891} {arXiv:1709.04891 [hep-th]} \BibitemShut
  {NoStop}%
\bibitem [{\citenamefont {Durieux}\ \emph
  {et~al.}(2020{\natexlab{a}})\citenamefont {Durieux}, \citenamefont
  {Kitahara}, \citenamefont {Shadmi},\ and\ \citenamefont
  {Weiss}}]{electroweak}%
  \BibitemOpen
  \bibfield  {author} {\bibinfo {author} {\bibfnamefont {G.}~\bibnamefont
  {Durieux}}, \bibinfo {author} {\bibfnamefont {T.}~\bibnamefont {Kitahara}},
  \bibinfo {author} {\bibfnamefont {Y.}~\bibnamefont {Shadmi}},\ and\ \bibinfo
  {author} {\bibfnamefont {Y.}~\bibnamefont {Weiss}},\ }\href
  {https://doi.org/10.1007/JHEP01(2020)119} {\bibfield  {journal} {\bibinfo
  {journal} {JHEP}\ }\bibfield  {number} {\bibinfo  {number} { (01)},\ \bibinfo
  {pages} {119}},\ }\Eprint {https://arxiv.org/abs/1909.10551}
  {arXiv:1909.10551 [hep-ph]} \BibitemShut {NoStop}%
\bibitem [{\citenamefont {Chung}\ \emph {et~al.}(2018)\citenamefont {Chung},
  \citenamefont {Huang}, \citenamefont {Kim},\ and\ \citenamefont
  {Lee}}]{spinningblackholes}%
  \BibitemOpen
  \bibfield  {author} {\bibinfo {author} {\bibfnamefont {M.-Z.}\ \bibnamefont
  {Chung}}, \bibinfo {author} {\bibfnamefont {Y.-t.}\ \bibnamefont {Huang}},
  \bibinfo {author} {\bibfnamefont {J.-W.}\ \bibnamefont {Kim}},\ and\ \bibinfo
  {author} {\bibfnamefont {S.}~\bibnamefont {Lee}},\ }\href@noop {} {\bibinfo
  {title} {The simplest massive {S}-matrix: from minimal coupling to black
  holes}} (\bibinfo {year} {2018}),\ \Eprint {https://arxiv.org/abs/1812.08752}
  {arXiv:1812.08752 [hep-th]} \BibitemShut {NoStop}%
\bibitem [{\citenamefont {Durieux}\ \emph
  {et~al.}(2020{\natexlab{b}})\citenamefont {Durieux}, \citenamefont
  {Kitahara}, \citenamefont {Machado}, \citenamefont {Shadmi},\ and\
  \citenamefont {Weiss}}]{massivecontact}%
  \BibitemOpen
  \bibfield  {author} {\bibinfo {author} {\bibfnamefont {G.}~\bibnamefont
  {Durieux}}, \bibinfo {author} {\bibfnamefont {T.}~\bibnamefont {Kitahara}},
  \bibinfo {author} {\bibfnamefont {C.~S.}\ \bibnamefont {Machado}}, \bibinfo
  {author} {\bibfnamefont {Y.}~\bibnamefont {Shadmi}},\ and\ \bibinfo {author}
  {\bibfnamefont {Y.}~\bibnamefont {Weiss}},\ }\href
  {https://doi.org/10.1007/JHEP12(2020)175} {\bibfield  {journal} {\bibinfo
  {journal} {JHEP}\ }\bibfield  {number} {\bibinfo  {number} { (12)},\ \bibinfo
  {pages} {175}},\ }\Eprint {https://arxiv.org/abs/2008.09652}
  {arXiv:2008.09652 [hep-ph]} \BibitemShut {NoStop}%
\bibitem [{\citenamefont {Dong}\ \emph {et~al.}(2023)\citenamefont {Dong},
  \citenamefont {Ma},\ and\ \citenamefont {Shu}}]{dongshort}%
  \BibitemOpen
  \bibfield  {author} {\bibinfo {author} {\bibfnamefont {Z.-Y.}\ \bibnamefont
  {Dong}}, \bibinfo {author} {\bibfnamefont {T.}~\bibnamefont {Ma}},\ and\
  \bibinfo {author} {\bibfnamefont {J.}~\bibnamefont {Shu}},\ }\href
  {https://doi.org/10.1103/PhysRevD.107.L111901} {\bibfield  {journal}
  {\bibinfo  {journal} {Phys. Rev. D}\ }\textbf {\bibinfo {volume} {107}},\
  \bibinfo {pages} {L111901} (\bibinfo {year} {2023})},\ \Eprint
  {https://arxiv.org/abs/2103.15837} {arXiv:2103.15837 [hep-ph]} \BibitemShut
  {NoStop}%
\bibitem [{\citenamefont {Dong}\ \emph {et~al.}(2022)\citenamefont {Dong},
  \citenamefont {Ma}, \citenamefont {Shu},\ and\ \citenamefont {Zheng}}]{dong}%
  \BibitemOpen
  \bibfield  {author} {\bibinfo {author} {\bibfnamefont {Z.-Y.}\ \bibnamefont
  {Dong}}, \bibinfo {author} {\bibfnamefont {T.}~\bibnamefont {Ma}}, \bibinfo
  {author} {\bibfnamefont {J.}~\bibnamefont {Shu}},\ and\ \bibinfo {author}
  {\bibfnamefont {Y.-H.}\ \bibnamefont {Zheng}},\ }\href
  {https://doi.org/10.1103/PhysRevD.106.116010} {\bibfield  {journal} {\bibinfo
   {journal} {Phys. Rev. D}\ }\textbf {\bibinfo {volume} {106}},\ \bibinfo
  {pages} {116010} (\bibinfo {year} {2022})},\ \Eprint
  {https://arxiv.org/abs/2202.08350} {arXiv:2202.08350 [hep-ph]} \BibitemShut
  {NoStop}%
\bibitem [{\citenamefont {De~Angelis}(2022)}]{deangelis}%
  \BibitemOpen
  \bibfield  {author} {\bibinfo {author} {\bibfnamefont {S.}~\bibnamefont
  {De~Angelis}},\ }\href {https://doi.org/10.1007/JHEP08(2022)299} {\bibfield
  {journal} {\bibinfo  {journal} {JHEP}\ }\bibfield  {number} {\bibinfo
  {number} { (08)},\ \bibinfo {pages} {299}},\ }\Eprint
  {https://arxiv.org/abs/2202.02681} {arXiv:2202.02681 [hep-th]} \BibitemShut
  {NoStop}%
\bibitem [{\citenamefont {Li}\ \emph {et~al.}(2021{\natexlab{a}})\citenamefont
  {Li}, \citenamefont {Ren}, \citenamefont {Shu}, \citenamefont {Xiao},
  \citenamefont {Yu},\ and\ \citenamefont {Zheng}}]{ybasiseight}%
  \BibitemOpen
  \bibfield  {author} {\bibinfo {author} {\bibfnamefont {H.-L.}\ \bibnamefont
  {Li}}, \bibinfo {author} {\bibfnamefont {Z.}~\bibnamefont {Ren}}, \bibinfo
  {author} {\bibfnamefont {J.}~\bibnamefont {Shu}}, \bibinfo {author}
  {\bibfnamefont {M.-L.}\ \bibnamefont {Xiao}}, \bibinfo {author}
  {\bibfnamefont {J.-H.}\ \bibnamefont {Yu}},\ and\ \bibinfo {author}
  {\bibfnamefont {Y.-H.}\ \bibnamefont {Zheng}},\ }\href
  {https://doi.org/10.1103/PhysRevD.104.015026} {\bibfield  {journal} {\bibinfo
   {journal} {Phys. Rev. D}\ }\textbf {\bibinfo {volume} {104}},\ \bibinfo
  {pages} {015026} (\bibinfo {year} {2021}{\natexlab{a}})},\ \Eprint
  {https://arxiv.org/abs/2005.00008} {arXiv:2005.00008 [hep-ph]} \BibitemShut
  {NoStop}%
\bibitem [{\citenamefont {Li}\ \emph {et~al.}(2021{\natexlab{b}})\citenamefont
  {Li}, \citenamefont {Ren}, \citenamefont {Xiao}, \citenamefont {Yu},\ and\
  \citenamefont {Zheng}}]{ybasisnine}%
  \BibitemOpen
  \bibfield  {author} {\bibinfo {author} {\bibfnamefont {H.-L.}\ \bibnamefont
  {Li}}, \bibinfo {author} {\bibfnamefont {Z.}~\bibnamefont {Ren}}, \bibinfo
  {author} {\bibfnamefont {M.-L.}\ \bibnamefont {Xiao}}, \bibinfo {author}
  {\bibfnamefont {J.-H.}\ \bibnamefont {Yu}},\ and\ \bibinfo {author}
  {\bibfnamefont {Y.-H.}\ \bibnamefont {Zheng}},\ }\href
  {https://doi.org/10.1103/PhysRevD.104.015025} {\bibfield  {journal} {\bibinfo
   {journal} {Phys. Rev. D}\ }\textbf {\bibinfo {volume} {104}},\ \bibinfo
  {pages} {015025} (\bibinfo {year} {2021}{\natexlab{b}})},\ \Eprint
  {https://arxiv.org/abs/2007.07899} {arXiv:2007.07899 [hep-ph]} \BibitemShut
  {NoStop}%
\bibitem [{\citenamefont {Li}\ \emph {et~al.}(2022)\citenamefont {Li},
  \citenamefont {Ren}, \citenamefont {Xiao}, \citenamefont {Yu},\ and\
  \citenamefont {Zheng}}]{li}%
  \BibitemOpen
  \bibfield  {author} {\bibinfo {author} {\bibfnamefont {H.-L.}\ \bibnamefont
  {Li}}, \bibinfo {author} {\bibfnamefont {Z.}~\bibnamefont {Ren}}, \bibinfo
  {author} {\bibfnamefont {M.-L.}\ \bibnamefont {Xiao}}, \bibinfo {author}
  {\bibfnamefont {J.-H.}\ \bibnamefont {Yu}},\ and\ \bibinfo {author}
  {\bibfnamefont {Y.-H.}\ \bibnamefont {Zheng}},\ }\href
  {https://doi.org/10.1007/JHEP04(2022)140} {\bibfield  {journal} {\bibinfo
  {journal} {JHEP}\ }\textbf {\bibinfo {volume} {04}}\bibfield  {number}
  {\bibinfo  {number} { (04)},\ \bibinfo {pages} {140}},\ }\Eprint
  {https://arxiv.org/abs/2201.04639} {arXiv:2201.04639 [hep-ph]} \BibitemShut
  {NoStop}%
\bibitem [{\citenamefont {El~Maazouz}\ \emph {et~al.}(2024)\citenamefont
  {El~Maazouz}, \citenamefont {Pfister},\ and\ \citenamefont
  {Sturmfels}}]{varieties}%
  \BibitemOpen
  \bibfield  {author} {\bibinfo {author} {\bibfnamefont {Y.}~\bibnamefont
  {El~Maazouz}}, \bibinfo {author} {\bibfnamefont {A.}~\bibnamefont
  {Pfister}},\ and\ \bibinfo {author} {\bibfnamefont {B.}~\bibnamefont
  {Sturmfels}},\ }\href@noop {} {\bibinfo {title} {Spinor-helicity varieties}}
  (\bibinfo {year} {2024}),\ \Eprint {https://arxiv.org/abs/2406.17331}
  {arXiv:2406.17331 [math.AG]} \BibitemShut {NoStop}%
\bibitem [{\citenamefont {Dong}\ \emph {et~al.}(2025)\citenamefont {Dong},
  \citenamefont {Ma}, \citenamefont {Yang},\ and\ \citenamefont
  {Zhou}}]{darkbasis}%
  \BibitemOpen
  \bibfield  {author} {\bibinfo {author} {\bibfnamefont {Z.}~\bibnamefont
  {Dong}}, \bibinfo {author} {\bibfnamefont {T.}~\bibnamefont {Ma}}, \bibinfo
  {author} {\bibfnamefont {C.}~\bibnamefont {Yang}},\ and\ \bibinfo {author}
  {\bibfnamefont {Z.}~\bibnamefont {Zhou}},\ }\href
  {https://doi.org/10.1007/JHEP07(2025)104} {\bibfield  {journal} {\bibinfo
  {journal} {JHEP}\ }\bibfield  {number} {\bibinfo  {number} { (07)},\ \bibinfo
  {pages} {104}},\ }\Eprint {https://arxiv.org/abs/2412.20096}
  {arXiv:2412.20096 [hep-ph]} \BibitemShut {NoStop}%
\bibitem [{\citenamefont {Chala}\ \emph {et~al.}(2024)\citenamefont {Chala},
  \citenamefont {L\'opez~Miras}, \citenamefont {Santiago},\ and\ \citenamefont
  {Vilches}}]{efficientmatching}%
  \BibitemOpen
  \bibfield  {author} {\bibinfo {author} {\bibfnamefont {M.}~\bibnamefont
  {Chala}}, \bibinfo {author} {\bibfnamefont {J.}~\bibnamefont
  {L\'opez~Miras}}, \bibinfo {author} {\bibfnamefont {J.}~\bibnamefont
  {Santiago}},\ and\ \bibinfo {author} {\bibfnamefont {F.}~\bibnamefont
  {Vilches}},\ }\href@noop {} {\bibinfo {title} {Efficient on-shell matching}}
  (\bibinfo {year} {2024}),\ \Eprint {https://arxiv.org/abs/2411.12798}
  {arXiv:2411.12798 [hep-ph]} \BibitemShut {NoStop}%
\bibitem [{\citenamefont {L\'opez~Miras}\ and\ \citenamefont
  {Vilches}(2025)}]{mosca}%
  \BibitemOpen
  \bibfield  {author} {\bibinfo {author} {\bibfnamefont {J.}~\bibnamefont
  {L\'opez~Miras}}\ and\ \bibinfo {author} {\bibfnamefont {F.}~\bibnamefont
  {Vilches}},\ }\href@noop {} {\bibinfo {title} {Automation of a matching
  on-shell calculator}} (\bibinfo {year} {2025}),\ \Eprint
  {https://arxiv.org/abs/2505.21353} {arXiv:2505.21353 [hep-ph]} \BibitemShut
  {NoStop}%
\bibitem [{\citenamefont {De~Laurentis}\ and\ \citenamefont
  {Page}(2022)}]{delaurentis}%
  \BibitemOpen
  \bibfield  {author} {\bibinfo {author} {\bibfnamefont {G.}~\bibnamefont
  {De~Laurentis}}\ and\ \bibinfo {author} {\bibfnamefont {B.}~\bibnamefont
  {Page}},\ }\href {https://doi.org/10.1007/JHEP12(2022)140} {\bibfield
  {journal} {\bibinfo  {journal} {JHEP}\ }\textbf {\bibinfo {volume}
  {12}}\bibfield  {number} {\bibinfo  {number} { (12)},\ \bibinfo {pages}
  {140}},\ }\Eprint {https://arxiv.org/abs/2203.04269} {arXiv:2203.04269
  [hep-th]} \BibitemShut {NoStop}%
\bibitem [{\citenamefont {Campbell}\ \emph {et~al.}(2024)\citenamefont
  {Campbell}, \citenamefont {De~Laurentis},\ and\ \citenamefont
  {Ellis}}]{hh3j}%
  \BibitemOpen
  \bibfield  {author} {\bibinfo {author} {\bibfnamefont {J.~M.}\ \bibnamefont
  {Campbell}}, \bibinfo {author} {\bibfnamefont {G.}~\bibnamefont
  {De~Laurentis}},\ and\ \bibinfo {author} {\bibfnamefont {R.~K.}\ \bibnamefont
  {Ellis}},\ }\href {https://doi.org/10.1007/JHEP10(2024)230} {\bibfield
  {journal} {\bibinfo  {journal} {JHEP}\ }\textbf {\bibinfo {volume}
  {10}}\bibfield  {number} {\bibinfo  {number} { (10)},\ \bibinfo {pages}
  {230}},\ }\Eprint {https://arxiv.org/abs/2408.12686} {arXiv:2408.12686
  [hep-ph]} \BibitemShut {NoStop}%
\bibitem [{\citenamefont {Campbell}\ \emph {et~al.}(2026)\citenamefont
  {Campbell}, \citenamefont {De~Laurentis},\ and\ \citenamefont {Ellis}}]{hhh}%
  \BibitemOpen
  \bibfield  {author} {\bibinfo {author} {\bibfnamefont {J.~M.}\ \bibnamefont
  {Campbell}}, \bibinfo {author} {\bibfnamefont {G.}~\bibnamefont
  {De~Laurentis}},\ and\ \bibinfo {author} {\bibfnamefont {R.~K.}\ \bibnamefont
  {Ellis}},\ }\href@noop {} {\bibfield  {journal} {\bibinfo  {journal} {JHEP}\
  }\textbf {\bibinfo {volume} {01}}\bibfield  {number} {\bibinfo  {number} {
  (01)},\ \bibinfo {pages} {157}},\ }\Eprint {https://arxiv.org/abs/2507.19313}
  {arXiv:2507.19313 [hep-ph]} \BibitemShut {NoStop}%
\bibitem [{\citenamefont {Campbell}\ \emph {et~al.}(2025)\citenamefont
  {Campbell}, \citenamefont {De~Laurentis},\ and\ \citenamefont
  {Ellis}}]{qttH}%
  \BibitemOpen
  \bibfield  {author} {\bibinfo {author} {\bibfnamefont {J.~M.}\ \bibnamefont
  {Campbell}}, \bibinfo {author} {\bibfnamefont {G.}~\bibnamefont
  {De~Laurentis}},\ and\ \bibinfo {author} {\bibfnamefont {R.~K.}\ \bibnamefont
  {Ellis}},\ }\href {https://doi.org/10.1007/JHEP07(2025)147} {\bibfield
  {journal} {\bibinfo  {journal} {JHEP}\ }\textbf {\bibinfo {volume}
  {07}}\bibfield  {number} {\bibinfo  {number} { (07)},\ \bibinfo {pages}
  {147}},\ }\Eprint {https://arxiv.org/abs/2504.19909} {arXiv:2504.19909
  [hep-ph]} \BibitemShut {NoStop}%
\bibitem [{\citenamefont {Cox}\ \emph {et~al.}(2015)\citenamefont {Cox},
  \citenamefont {Little},\ and\ \citenamefont {O'Shea}}]{coxLittleOShea}%
  \BibitemOpen
  \bibfield  {author} {\bibinfo {author} {\bibfnamefont {D.~A.}\ \bibnamefont
  {Cox}}, \bibinfo {author} {\bibfnamefont {J.}~\bibnamefont {Little}},\ and\
  \bibinfo {author} {\bibfnamefont {D.}~\bibnamefont {O'Shea}},\ }\href
  {https://doi.org/10.1007/978-3-319-16721-3} {\emph {\bibinfo {title} {Ideals,
  Varieties, and Algorithms}}},\ \bibinfo {edition} {4th}\ ed.\ (\bibinfo
  {publisher} {Springer},\ \bibinfo {address} {Cham},\ \bibinfo {year}
  {2015})\BibitemShut {NoStop}%
\bibitem [{\citenamefont {Henning}\ and\ \citenamefont
  {Melia}(2019)}]{henning}%
  \BibitemOpen
  \bibfield  {author} {\bibinfo {author} {\bibfnamefont {B.}~\bibnamefont
  {Henning}}\ and\ \bibinfo {author} {\bibfnamefont {T.}~\bibnamefont
  {Melia}},\ }\href {https://doi.org/10.1103/PhysRevD.100.016015} {\bibfield
  {journal} {\bibinfo  {journal} {Phys. Rev. D}\ }\textbf {\bibinfo {volume}
  {100}},\ \bibinfo {pages} {016015} (\bibinfo {year} {2019})},\ \Eprint
  {https://arxiv.org/abs/1902.06754} {arXiv:1902.06754 [hep-ph]} \BibitemShut
  {NoStop}%
\bibitem [{\citenamefont {Hagiwara}\ \emph {et~al.}(1987)\citenamefont
  {Hagiwara}, \citenamefont {Peccei}, \citenamefont {Zeppenfeld},\ and\
  \citenamefont {Hikasa}}]{hagiwara1987}%
  \BibitemOpen
  \bibfield  {author} {\bibinfo {author} {\bibfnamefont {K.}~\bibnamefont
  {Hagiwara}}, \bibinfo {author} {\bibfnamefont {R.~D.}\ \bibnamefont
  {Peccei}}, \bibinfo {author} {\bibfnamefont {D.}~\bibnamefont {Zeppenfeld}},\
  and\ \bibinfo {author} {\bibfnamefont {K.}~\bibnamefont {Hikasa}},\ }\href
  {https://doi.org/10.1016/0550-3213(87)90685-7} {\bibfield  {journal}
  {\bibinfo  {journal} {Nucl. Phys. B}\ }\textbf {\bibinfo {volume} {282}},\
  \bibinfo {pages} {253} (\bibinfo {year} {1987})}\BibitemShut {NoStop}%
\bibitem [{\citenamefont {Zeppenfeld}\ and\ \citenamefont
  {Willenbrock}(1988)}]{zeppenfeld1988}%
  \BibitemOpen
  \bibfield  {author} {\bibinfo {author} {\bibfnamefont {D.}~\bibnamefont
  {Zeppenfeld}}\ and\ \bibinfo {author} {\bibfnamefont {S.}~\bibnamefont
  {Willenbrock}},\ }\href {https://doi.org/10.1103/PhysRevD.37.1775} {\bibfield
   {journal} {\bibinfo  {journal} {Phys. Rev. D}\ }\textbf {\bibinfo {volume}
  {37}},\ \bibinfo {pages} {1775} (\bibinfo {year} {1988})}\BibitemShut
  {NoStop}%
\bibitem [{\citenamefont {Zhou}()}]{massiveamplitudecode}%
  \BibitemOpen
  \bibfield  {author} {\bibinfo {author} {\bibfnamefont {Z.-Z.}\ \bibnamefont
  {Zhou}},\ }\href@noop {} {\bibinfo {title} {{MassiveAmplitude}}},\ \bibinfo
  {howpublished} {\url{https://github.com/zizhengzhou/MassiveAmplitude}},\
  \bibinfo {note} {revision a84664d, accessed October 9, 2026}\BibitemShut
  {NoStop}%
\end{thebibliography}%

\newpage

\renewcommand{\theequation}{E\arabic{equation}}
\section*{End Matter}

\subsection*{Tensor contraction ordering}
The full monomial order compares bracket degree, the unbolded bracket product, and finally the contraction structure. Here we define the third comparison explicitly. A tensor monomial $T$ decomposes into a tensor product of connected contraction factors, $T=t_1\otimes t_2\otimes \cdots\otimes t_m$. First, we compare the factor \textit{lengths} $\ell(t_i)$, defined as the number of constituent spinor brackets in each connected factor. Sorting the lengths in descending order, $\ell(t_1)\ge \ell(t_2)\ge\cdots$, we compare the length vectors $(\ell(t_1),\ell(t_2),\ldots)$ lexicographically; for example,
\begin{equation}
	\operatorname{tr}(p_2\bar p_3p_2\bar p_4)
	\succ \operatorname{tr}(p_2\bar p_3)
	\operatorname{tr}(p_2\bar p_4),
\end{equation}
because $(4) >_{\rm lex} (2,2)$. For single factors of equal length, open chains precede closed traces, \textit{i.e.}, $\pp{\boldsymbol1}{2}{\boldsymbol1} \succ \operatorname{tr}(p_1\bar p_2)$.

When both comparisons tie, we compare the unbolded bracket products of the individual connected factors using the massless order~\eqref{eq:variable-order}, starting from the longest factor. Consider the tensors in Eq.~\eqref{eq:tensor_T} of the main text,
\begin{align}
	T_1^{(r)}&=\an{\boldsymbol1}{\boldsymbol n}
	\pp{\boldsymbol2}{n}{\boldsymbol r},&
	T_2^{(r)}&=\an{\boldsymbol2}{\boldsymbol n}
	\pp{\boldsymbol1}{n}{\boldsymbol r}.
\end{align}
Both share the overall unbolded product $\an{1}{n}\an{2}{n}\sq{n}{r}$ and have the length vector $(2,1)$. Comparing their length-two factors gives $\an{2}{n}\sq{n}{r}\succ\an{1}{n}\sq{n}{r}$, and hence $T_1^{(r)}\succ T_2^{(r)}$. If multiple connected factors share the same length, we sort those factors in increasing order (putting closed traces before open chains) and compare the sorted lists from left to right at their first difference. Attaching $\sq{\boldsymbol{n}}{\boldsymbol{s}}$ to $T_1^{(r)}$ and $T_2^{(r)}$ yields the $(2,2)$ tensors of the main text,
\begin{align}
	D_1^{(r,s)}&=\pp{\boldsymbol1}{n}{\boldsymbol s}
	\pp{\boldsymbol2}{n}{\boldsymbol r},&
	D_2^{(r,s)}&=\pp{\boldsymbol2}{n}{\boldsymbol s}
	\pp{\boldsymbol1}{n}{\boldsymbol r}.
\end{align}
Take $2<r<s<n$. Since $\sq{n}{s}\prec\sq{n}{r}$ under Eq.~\eqref{eq:variable-order}, the chain ending at $s$ is the smaller factor and comes first in each sorted list. Comparing those first chains gives $\an{1}{n}\sq{n}{s}\prec\an{2}{n}\sq{n}{s}$, and hence $D_1^{(r,s)}\prec D_2^{(r,s)}$, illustrating how contracting with a context can reverse the relative order of two tensors.

Any remaining tie is resolved by the particle-label sequence along each factor, followed by its angle/square chirality sequence. For example, among open chains with the same bracket content, a canonical sequence beginning $(1,2,\ldots)$ precedes one beginning $(1,3,\ldots)$. We choose the smaller endpoint-to-endpoint orientation of an open chain and the lexicographically smallest cyclic or reversed presentation of a trace, keeping associated signs in the coefficient. These final comparisons distinguish contraction routings without assigning an order to dummy names of free indices.

\subsection*{Rule catalog construction and the $n=3$ example}
For general $n$, the degree-two starting pool consists of the $n^2$ momentum identities $\mathsf M_{ij}$, the $2\binom{n}{4}$ distinct-label Schouten identities $\mathsf S^{\rm a,s}_{ijkl}$, and the repeated-label Schouten identities---$2n\binom{n-1}{2}$ of type $\mathsf S^{\rm a,s}_{iijk}$ and $2\binom{n}{2}$ of type $\mathsf S^{\rm a,s}_{iijj}$---giving $N_{d=2}^{\rm orig}(n)=21,54,115$ original rules for $n=3,4,5$. Mass-determinant relations are imposed prior to reduction and are not counted separately. At three points, the 21 originals comprise nine momentum identities $\mathsf M_{ij}$ and twelve repeated-label Schouten identities ($\mathsf S^{\rm a,s}_{iijj}$ and $\mathsf S^{\rm a,s}_{iijk}$). Cancelling actual contextual LMs in minimal common contractions of these originals generates six further rules:
\begin{center}\small
	\begin{tabular}{ccl}
		\toprule
		& Generating pair & Common leading monomial\\
		\midrule
		$S_0$ & $(\mathsf M_{11},\mathsf M_{22})$ & $s_{12}$\\
		$S_1$ & $(\mathsf S^{\rm a}_{1223},\mathsf M_{31})$ & $\an{\boldsymbol 1}{\boldsymbol 2}[\boldsymbol1|p_2|\boldsymbol3\rangle$\\
		$S_2$ & $(\mathsf S^{\rm s}_{1123},\mathsf M_{22})$ & $\sq{\boldsymbol 1}{\boldsymbol 3}[\boldsymbol 2|p_{1}|\boldsymbol 2\rangle$\\
		$S_3$ & $(\mathsf S^{\rm s}_{1123},\mathsf M_{32})$ & $\sq{\boldsymbol 1}{\boldsymbol 3}[\boldsymbol2|p_1|\boldsymbol3\rangle$\\
		$S_4$ & $(\mathsf S^{\rm a}_{1233},\mathsf S^{\rm s}_{1233})$ & $\an{\boldsymbol 3}{\boldsymbol 1}\pp{\boldsymbol 2}{3}{\boldsymbol 2}\sq{\boldsymbol 3}{\boldsymbol 1}$\\
		$S_5$ & $(\mathsf S^{\rm s}_{1223},\mathsf S^{\rm a}_{1223})$ & $\an{\boldsymbol 2}{\boldsymbol 3}[\boldsymbol2|p_1\bar p_2|\boldsymbol3]$\\
		\bottomrule
	\end{tabular}
\end{center}
For $S_0$, the scalar contractions $C_i=2p_i\cdot(p_1+p_2+p_3)=0$ of $\mathsf M_{ii}$ give $C_1=s_{12}+s_{13}+2m_1^2$ and $C_2=s_{12}+s_{23}+2m_2^2$, which share the LM $s_{12}$; their difference gives $S_0=C_1-C_2$. Normalizing the remaining cancellations by overall nonzero mass factors and signs (with $S_4$ as in the main text) gives the unreduced $S$-polynomials
\begin{align}
	S_0={}&s_{13}-s_{23}+2m_1^2-2m_2^2,\\
	S_1={}&\an{\boldsymbol 2}{\boldsymbol 3}[\boldsymbol 1|p_{2}|\boldsymbol 1\rangle+m_1\an{\boldsymbol 1}{\boldsymbol 2}\an{\boldsymbol 1}{\boldsymbol 3}\nonumber\\[-2pt]
	&+m_2\sq{\boldsymbol 1}{\boldsymbol 2}\an{\boldsymbol 1}{\boldsymbol 3}+m_3\an{\boldsymbol 1}{\boldsymbol 2}\sq{\boldsymbol 1}{\boldsymbol 3},\\
	S_2={}&\sq{\boldsymbol 1}{\boldsymbol 2}\pp{\boldsymbol2}{1}{\boldsymbol3}-\sq{\boldsymbol 1}{\boldsymbol 3}[\boldsymbol 2|p_{3}|\boldsymbol 2\rangle+m_1\an{\boldsymbol 1}{\boldsymbol 2}\sq{\boldsymbol 2}{\boldsymbol 3},\\
	S_3={}&\sq{\boldsymbol 1}{\boldsymbol 2}[\boldsymbol 3|p_{1}|\boldsymbol 3\rangle-m_2\sq{\boldsymbol 1}{\boldsymbol 3}\an{\boldsymbol 2}{\boldsymbol 3}\nonumber\\[-2pt]
	&-m_1\an{\boldsymbol 1}{\boldsymbol 3}\sq{\boldsymbol 2}{\boldsymbol 3}-m_3\sq{\boldsymbol 1}{\boldsymbol 3}\sq{\boldsymbol 2}{\boldsymbol 3},\\
	S_4={}&\an{\boldsymbol 2}{\boldsymbol 3}\pp{\boldsymbol 1}{3}{\boldsymbol 2}\sq{\boldsymbol 3}{\boldsymbol 1}-\an{\boldsymbol 3}{\boldsymbol 1}\pp{\boldsymbol 2}{3}{\boldsymbol 1}\sq{\boldsymbol 2}{\boldsymbol 3}\nonumber\\[-2pt]
	&+m_3\sq{\boldsymbol 1}{\boldsymbol 2}\an{\boldsymbol 3}{\boldsymbol 1}\an{\boldsymbol 2}{\boldsymbol 3}+m_3\an{\boldsymbol 1}{\boldsymbol 2}\sq{\boldsymbol 3}{\boldsymbol 1}\sq{\boldsymbol 2}{\boldsymbol 3},\\
	S_5={}&[\boldsymbol 2|p_{1}|\boldsymbol 2\rangle[\boldsymbol 3|p_{2}|\boldsymbol 3\rangle-\sq{\boldsymbol 2}{\boldsymbol 3}\an{\boldsymbol 2}{\boldsymbol 3}s_{12}\nonumber\\[-2pt]
	&-m_2\an{\boldsymbol 2}{\boldsymbol 3}\pp{\boldsymbol2}{1}{\boldsymbol3}-m_2\sq{\boldsymbol 2}{\boldsymbol 3}[\boldsymbol2|p_1|\boldsymbol3\rangle.
\end{align}
We now reduce each $S_i$ using the established identities. For $S_0$, eliminating $s_{13}$ via $C_3=s_{13}+s_{23}+2m_3^2=0$ yields $S_0\to -2R_0$\footnote{The same cancellation produces the single generated $d=2$ rule at arbitrary $n$, giving the $d=2$ totals $22,55,116$ in Table~\ref{tab:rule-degree-counts}.}. Meanwhile, $S_1,\ldots,S_5$ reduce to $-R_1,\ldots,-R_5$ in $1,1,1,2,6$ steps, respectively. Displayed with the LM first, the six resulting rules $R_i=0$ are
\begin{align}
	R_0={}&s_{23}-m_{1}^2+m_{2}^2+m_{3}^2,\\
	R_1={}&\an{\boldsymbol{2}}{\boldsymbol{3}}[\boldsymbol{1}|p_{3}|\boldsymbol{1}\rangle-m_{1}\an{\boldsymbol{1}}{\boldsymbol{2}}\an{\boldsymbol{1}}{\boldsymbol{3}}\nonumber\\[-2pt]
	&-m_{2}\sq{\boldsymbol{1}}{\boldsymbol{2}}\an{\boldsymbol{1}}{\boldsymbol{3}}-m_{3}\an{\boldsymbol{1}}{\boldsymbol{2}}\sq{\boldsymbol{1}}{\boldsymbol{3}},\\
	R_2={}&\sq{\boldsymbol{1}}{\boldsymbol{3}}[\boldsymbol{2}|p_{3}|\boldsymbol{2}\rangle-m_{3}\sq{\boldsymbol{1}}{\boldsymbol{2}}\an{\boldsymbol{2}}{\boldsymbol{3}}\nonumber\\[-2pt]
	&-m_{1}\an{\boldsymbol{1}}{\boldsymbol{2}}\sq{\boldsymbol{2}}{\boldsymbol{3}}-m_{2}\sq{\boldsymbol{1}}{\boldsymbol{2}}\sq{\boldsymbol{2}}{\boldsymbol{3}},\\
	R_3={}&\sq{\boldsymbol{1}}{\boldsymbol{2}}[\boldsymbol{3}|p_{2}|\boldsymbol{3}\rangle+m_{2}\sq{\boldsymbol{1}}{\boldsymbol{3}}\an{\boldsymbol{2}}{\boldsymbol{3}}\nonumber\\[-2pt]
	&+m_{1}\an{\boldsymbol{1}}{\boldsymbol{3}}\sq{\boldsymbol{2}}{\boldsymbol{3}}+m_{3}\sq{\boldsymbol{1}}{\boldsymbol{3}}\sq{\boldsymbol{2}}{\boldsymbol{3}},\\
	R_4={}&m_{3}\sq{\boldsymbol{1}}{\boldsymbol{2}}\an{\boldsymbol{1}}{\boldsymbol{3}}\an{\boldsymbol{2}}{\boldsymbol{3}}+m_{2}\an{\boldsymbol{1}}{\boldsymbol{2}}\sq{\boldsymbol{1}}{\boldsymbol{3}}\an{\boldsymbol{2}}{\boldsymbol{3}}\nonumber\\[-2pt]
	&+m_{1}\sq{\boldsymbol{1}}{\boldsymbol{2}}\sq{\boldsymbol{1}}{\boldsymbol{3}}\an{\boldsymbol{2}}{\boldsymbol{3}}+m_{1}\an{\boldsymbol{1}}{\boldsymbol{2}}\an{\boldsymbol{1}}{\boldsymbol{3}}\sq{\boldsymbol{2}}{\boldsymbol{3}}\nonumber\\[-2pt]
	&+m_{2}\sq{\boldsymbol{1}}{\boldsymbol{2}}\an{\boldsymbol{1}}{\boldsymbol{3}}\sq{\boldsymbol{2}}{\boldsymbol{3}}+m_{3}\an{\boldsymbol{1}}{\boldsymbol{2}}\sq{\boldsymbol{1}}{\boldsymbol{3}}\sq{\boldsymbol{2}}{\boldsymbol{3}},\\
	R_5={}&[\boldsymbol{2}|p_{3}|\boldsymbol{2}\rangle[\boldsymbol{3}|p_{2}|\boldsymbol{3}\rangle+m_{2}m_{3}\left(\an{\boldsymbol{2}}{\boldsymbol{3}}\an{\boldsymbol{2}}{\boldsymbol{3}}+\sq{\boldsymbol{2}}{\boldsymbol{3}}\sq{\boldsymbol{2}}{\boldsymbol{3}}\right)\nonumber\\[-2pt]
	&+(-m_{1}^2+m_{2}^2+m_{3}^2)\sq{\boldsymbol{2}}{\boldsymbol{3}}\an{\boldsymbol{2}}{\boldsymbol{3}},
\end{align}
where $R_4=-\mathsf F$ is the cubic three-vector identity of Eq.~\eqref{eq:three-vector-F} (with $\an{\boldsymbol 1}{\boldsymbol 3}=-\an{\boldsymbol 3}{\boldsymbol 1}$ and $\sq{\boldsymbol 1}{\boldsymbol 3}=-\sq{\boldsymbol 3}{\boldsymbol 1}$). These six additions have LM degrees $2,3,3,3,3,4$, giving $21+6=27$ retained rules; all higher-degree overlaps at $n=3$ reduce to zero.

In general, the overlap search can be pruned by a tensor version of Buchberger's product criterion: if $H=\mathcal C(L,M)$ is formed from disjoint bracket factors of two monic rules $g=L-A$ and $h=M-B$ under a bilinear contraction map $\mathcal C$, their $S$-polynomial is
\begin{equation}
	S_H=\mathcal C(g,M)-\mathcal C(L,h)=\mathcal C(L,B)-\mathcal C(A,M).
\end{equation}
Whenever $L\to A$ inside $\mathcal C(L,B)$ and $M\to B$ inside $\mathcal C(A,M)$ remain valid contextual reductions on every term, both reduce to $\mathcal C(A,B)$ and $S_H\to 0$, so the pair can be skipped (whereas in $S_4$ above, the contraction across leg~3 creates momentum chains that intercept the reduction and yield $R_4\ne 0$). To bound the search in constructing a catalog $(n,d)$, the cutoff $d$ restricts the bracket degree of the \emph{retained remainder's LM}, $\deg(\LM(R_i))\le d$, rather than the degree of the intermediate common monomial, since mass determinants allow a degree-$(d+1)$ overlap to cancel down to a degree-$d$ rule (as in $S_4\to -R_4$). At $n=4$, the $(4,6)$ catalog contains 179 rules, with no new rules found through $d=8$; at $n=5$, the $(5,6)$ catalog contains 1,077 rules (retaining LMs through $d=6$ from common overlaps through degree seven).

\subsection*{Four-vector basis validation and code usage}
To validate completeness in the four-vector sector ($V^4D^{0,2,4}$), we enumerate all $39,945$ Lorentz-covariant candidate monomials with two symmetrized spinor indices on each vector leg. For massive spin-one wavefunctions, the spacetime derivative order is
\begin{equation}
	D = q + n_F,
\end{equation}
where $q$ counts explicit momentum insertions $p_k$ and $n_F$ counts same-chirality free spinor pairs on the same leg representing field strengths $F_{\mu\nu}$ (corresponding to effective operators of dimension $4+D$). Including both parities, reducing all candidates leaves $4$, $56$, and $141$ irreducible monomials at $D=0,2,4$ (cumulative dimensions $4, 60, 201$). Evaluating the $201$ irreducible tensors at generic rational on-shell kinematic points over a finite field $\mathbb F_p$ yields a $201\times 201$ matrix of full rank, certifying generic linear independence without floating-point error. Independent on-shell evaluations likewise confirm that each benchmark normal form in Table~\ref{tab:implementation-benchmarks} matches its input amplitude identically.

At each descending bracket degree, amplitude reduction applies rules in two stages: Stage~1 exhausts replacements that lower the bracket degree or decrease the unbolded bracket product, and Stage~2 then applies replacements that only reduce the contraction structure at fixed unbolded product. Because a contraction context can modify the right-hand side (RHS) of a matched rule, the algorithm holds the matched contextual LM fixed and reduces its complete RHS using established rules before classifying the replacement into Stage~1 or Stage~2. Once prepared, this contextual replacement (retaining the matched monomial and its contraction context) is cached so that subsequent appearances of the same contextual structure are reduced in a single step.

The \href{https://github.com/XiaoMing-lei/MassiveAmplitudeReduction}{public code repository} supplies the rule catalogs $(3,4)$, $(4,6)$, and $(5,6)$, Python code for catalog construction, and a Mathematica notebook illustrating amplitude input, reduction, and trace inspection. Using \texttt{ab}, \texttt{sb}, \texttt{m}, and \texttt{s} for angle brackets, square brackets, masses, and $s_{ij}=2p_i\cdot p_j$, the benchmark $\mathcal A_1$ is reduced via
\begin{verbatim}
	amp = ab[1,2] sb[2,3] sb[3,4] *
          SpinorChain[{1,3,2,4},1];
	result = AmplitudeReduce[amp];
	AmplitudeForm[result]
\end{verbatim}
Setting \texttt{ReductionTrace -> True} records the derivation history for inspection with \texttt{ReductionSteps}, displaying the $N_{\rm step}=14$ successive rule applications along the selected reduction sequence.

\end{document}